\documentclass[a4paper,11pt]{article}   
\usepackage{lineno,hyperref}
\usepackage[utf8]{inputenc}
\usepackage[english]{babel}
\usepackage{ifthen}
\usepackage{amssymb,amsmath}
\usepackage[usenames]{color}
\date{}

 \newif\ifNoRemark
    \def\addtheorem#1#2#3#4{ 
    \ifthenelse{\expandafter\isundefined\csname the#2\endcsname}{\newcounter{#2}}{}
    \newenvironment{#1}[1][\global\NoRemarktrue]
     {\par\addvspace{2mm}\noindent
       \refstepcounter{#2}{\bf #3~\csname the#2\endcsname
      \vphantom{##1}\ifNoRemark.\ \else\ (##1).\fi}\begingroup #4}%
     {\endgroup\par\addvspace{1mm}\global\NoRemarkfalse}
    \expandafter\newcommand\csname b#1\endcsname{\begin{#1}}
    \expandafter\newcommand\csname e#1\endcsname{\end{#1}}
    }

\addtheorem{theorem}{thrm}{Theorem}{\sl}
\addtheorem{lemma}{lmm}{Lemma}{\sl}
\addtheorem{proposition}{prpstn}{Proposition}{\sl}
\addtheorem{corollary}{crllr}{Corollary}{\sl}
\addtheorem{problem}{problem}{Problem}{\sl}
\addtheorem{remark}{}{Remark}{}

\begin{document}

\title{On  weight spectrum of linear codes
\thanks{The work was supported by the program of fundamental scientific
researches of the SB RAS  I.5.1, project No. 0314-2019-0017.}}

\author{ Vladimir N. Potapov\\ {\em \small Sobolev Institute of Mathematics, Novosibirsk,
Russia; email: vpotapov@math.nsc.ru}}

 \maketitle

\begin{abstract}
 We study sequences of linear or affine codes with uniform weight
 spectrum, i.e., a part of codewords with any fixed weight tends to zero.
 It is proved that a sequence of linear
codes has a uniform weight spectrum if the number  of vectors from
codes with weight $1$ grows to infinity. We find an example of a
sequence of linear codes such that the dimension of the code is the
half of the codelength but it has not a uniform weight
 spectrum.  This example generates eigenfunctions of the Fourier
transform with minimal support and partial covering sets. Moreover,
we generalize some MacWilliams-type identity.
\end{abstract}


\section{Introduction}

Let $Q_q^n$ be a vector space of  dimension $n$ over the Galois
field $GF(q)$.
 The weight
${\rm wt}(x)$ of a vector $x=(x_1,\dots,x_n)\in Q^n_q$ is the number
of nonzero coordinates $x_i$ of $x$. A support ${\rm supp}(x)$ of
$x\in Q^n_q$ is the set of nonzero coordinates of $x$. The
cardinality of the support is the weight of the vector, i.e., $|{\rm
supp}(x)|={\rm wt}(x)$. The support of a function $f$ is the set of
arguments $x$ such that $f(x)\neq 0$. Denote by ${\cal
A}_i(V)=\{x\in V : {\rm wt}(x)=i\}$ the subset of $V\subset Q_q^n$
which consists of the vectors with weight $i$. A code $V$ is called
$t$-weight if ${\cal A}_i(V)\neq \varnothing$ only for $t$ different
weights.
 A finite sequence  $|{\cal A}_i(V)|$, $i=0,\dots,n$,
is called the weight distribution of $V$. For any subset $V\subset
Q_q^n$ denote by $V^\bot=\{ u\in  Q_q^n : (u, v)=0\ {\rm for \ any
}\ v\in V\}$ the dual subspace. The finite sequence $|{\cal
A}_i(V^\bot)|$, $i=0,\dots,n$, is called a dual weight distribution
of $V$. We will use the  notation ${\cal A}_i={\cal A}_i(Q^n_q)$ for
brevity.

For a subspace (affine code) $V\subseteq Q_q^n$ we introduce the
notation $\alpha(V)=\max\limits_{i,x}\frac{|\mathcal{A}_i(V+
x)|}{|V|}$ where $x\in Q_q^n$. We will say that a sequence of
subspaces $V_n\subseteq Q^{m_n}_q$ has a uniform weight spectrum if
$\alpha(V_n)\rightarrow 0$ as $n\rightarrow\infty$. Let us note that
$\dim V_n\rightarrow \infty$ if a sequence  $V_n$ has a uniform
weight spectrum. It is clear that the sequence of spaces $Q^n_q$ has
a uniform spectrum and there exist sequences that consist of
$t$-weight codes, for example, Hadamard codes. Delsarte (see
Corollary \ref{corDelsarte}) proved that  the cardinality of a
$t$-weight linear code is upper-bounded by the polynomial  of degree
$t$ with respect to the codelength. It is natural to suggest that
the dimensions of codes from a nonuniform weight spectrum sequence
are significantly less than their  codelengthes.

Sidel'nikov \cite{Sid} posed a question about the difference between
the weight distribution spectrum of the whole vector space and the
weight distribution spectrum of a random linear code. Mathematical
expectation of the difference between the spectrum of random code
and the spectrum   of vector space (more precisely,  the spectrum
that is proportional to the spectrum of vector space) is estimated
in \cite{Leont} and \cite{Linial} for binary vector spaces. As
expected, a sequence of random linear codes has a uniform weight
spectrum. The weight distribution  of a code and one of the dual
code are connected by MacWilliams-type identities \cite{MS},
\cite{Pless}.
 In
\cite{Leont} it is presented a new MacWilliams-type identity
(\ref{eqLeont}) for spectra of linear binary codes. In this paper we
generalize this identity (Lemma \ref{lemLeo}). Moreover, we obtain
the following results. We prove that if $|{\cal
A}_1(C_n)|\rightarrow\infty$ then a sequence $C_n$ of linear codes
has a uniform weight spectrum. But a similar condition $|{\cal
A}_2(C_n)|\rightarrow\infty$ is not sufficient. We give estimations
for the cardinality of $t$-weight affine  codes (Corollaries
\ref{corAf1} and \ref{KrEigen}).
 We present an
example of a nonuniform weight spectrum sequence of binary linear
codes such that the dimension of the code is equal to one half of
the codelength. It is proved that this example is extremal in the
sense of the Fourier transform. It generates eigenfunctions of the
Fourier transform with  minimal supports.  We suggest that this
example is an extremal in the sense of covering arrays (Corollary
\ref{corAf5}).

\section{New identity on weight spectrum}

It is well known the  MacWilliams identity
\begin{equation}\label{eqMW}
|V^\perp| |{\cal A}_k({V})|=\sum\limits_m P_k(m;n,q)|{\cal
A}_m(V^\perp)|,
\end{equation}

where $V\subset Q^n_2$ is a subspace, $P_k(m;n,q)$ is the $q$-ary
Krawtchouk polynomial, i.e.,
\begin{equation}\label{eqKr}
P_k(m;n,q)= P_k(m)=
\sum\limits_{s=0}^k
(-1)^s(q-1)^{k-s}{{n-m}\choose{k-s}}{{m}\choose{s}},
\end{equation}
   if $m< s$ or
$ n-m<k-s$ then corresponding terms are equal to zero.

Consider a known generalization of (\ref{eqMW}) for functions
$f:Q^n_q\rightarrow \mathbb{C}$. Let us introduce the notations
$A_k[f]=\sum\limits_{{\rm wt}(z)=k}f(x)$ and
$\widehat{A}_k[f]=\sum\limits_{{\rm wt}(z)=k}\widehat{f}(z)$ where
$\widehat{f}(z)=\frac{1}{q^{n/2}}\sum\limits_{x\in
Q^q_n}f(x)\left(e^{\frac{2\pi i}{q}}\right)^{(x,z)}$ are
coefficients of the Fourier transform. The functions
$\left(e^{\frac{2\pi i}{q}}\right)^{(x,z)}$ are known as characters
of group $(\mathbb{Z}_q)^n$. If $f$ is the indicator of the subspace
$V$ then $A_k[f]=|{\cal A}_k({V})|$ and $\widehat{A}_k[f]=q^{\dim V
-n/2}|{\cal A}_m(V^\perp)|$. It holds 
\begin{equation}\label{c:eqMW}
 \widehat{A}_k[f]=
\frac{1}{q^{n/2}}\sum\limits_{m=0}^n P_k(m;n,q)A_m[f].
\end{equation}
From the another MacWilliams identity
\begin{equation}\label{eqMW2}
q^{n/2}\sum\limits_{k=0}^n\widehat{A}_k[f]z^k=\sum\limits_{m=0}^n
A_m[f](1-z)^m(1+(q-1)z)^{n-m}
\end{equation}
we obtain a generating function  for the coefficients of $q$-ary
Krawtchouk polynomial $P_k(m;n,q)$
\begin{equation}\label{eqMW4}
\sum\limits_{k=0}^nP_k(m;n,q)z^k= (1-z)^m(1+(q-1)z)^{n-m}
\end{equation}
and equation (\ref{eqKr}). We use (\ref{eqMW4}) to prove the
following identity.

\begin{proposition}\label{propKr}
$$\sum\limits_{m=0}^nP_k(m;n,q)=\frac{1}{q}{n+1 \choose
k+1}((q-1)^{k+1}-(-1)^{k+1}).$$
\end{proposition}
Proof.
$$\sum\limits_{m=0}^n\sum\limits_{k=0}^nP_k(m;n,q)z^k=(1+(q-1)z)^{n}
\sum\limits_{m=0}^n\left(\frac{1-z}{1+(q-1)z}\right)^m=$$
$$(1+(q-1)z)^{n}\left(\left(\frac{1-z}{1+(q-1)z}\right)^{n+1}-1\right)/
\left(\frac{1-z}{1+(q-1)z}-1\right)=$$
$$\frac{(1-z)^{n+1}-(1+(q-1)z)^{n+1}}{(1-z)-(1+(q-1)z)}=
\frac{1}{q}\sum\limits_{k=0}^{n} z^k((q-1)^{k+1}-(-1)^{k+1}){n+1
\choose k+1}.\quad \square$$

By double counting the values of the character $\left(e^{\frac{2\pi
i}{q}}\right)^{(u,v)}$ where $u\in {\cal A}_m(Q^n_q)$ and $v\in
{\cal A}_k(Q^n_q)$, we obtain another known equation
\begin{equation}\label{eqMW3}
P_k(m;n,q)|{\cal A}_m(Q^n_q)|=P_m(k;n,q)|{\cal A}_k(Q^n_q)|.
\end{equation}

\begin{lemma}\label{lemLeo}
$$\sum\limits_{k=0}^n\frac{\widehat{A}_k[f]}{(q-1)^k{n \choose k}}=
\frac{(n+1)(q-1)}{q^{1+n/2}}\sum\limits_{m=0}^n\frac{
A_m[f]}{m+1}\left(1-\frac{(-1)^{m+1}}{(q-1)^{m+1}}\right).$$
\end{lemma}
Proof. From (\ref{c:eqMW}) and (\ref{eqMW3}) we obtain
$$
\sum\limits_{k=0}^n\widehat{A}_k[f]=
\frac{1}{q^{n/2}}\sum\limits_{k=0}^n\sum\limits_{m=0}^n
P_m(k;n,q)\frac{|{\cal A}_k(Q^n_q)|A_m[f]}{|{\cal A}_m(Q^n_q)|}.$$
Therefore
$$
\sum\limits_{k=0}^n\frac{\widehat{A}_k[f]}{|{\cal A}_k(Q^n_q)|}=
\frac{1}{q^{n/2}}\sum\limits_{m=0}^n\left(\frac{A_m[f]}{|{\cal
A}_m(Q^n_q)|} \sum\limits_{k=0}^nP_m(k;n,q)\right).$$

By Proposition \ref{propKr} and equation $|{\cal
A}_k(Q^n_q)|=(q-1)^k{n \choose k}$ we obtain the required formula.
$\square$

If $f$ is the indicator of a subspace $V$ then we have the following

\begin{corollary}\label{corLeont0} Let $V\in Q^n_2$ be a subspace. It holds
$$\sum\limits_{k=0}^n\frac{|{\cal A}_k(V)|}{(q-1)^k{n \choose k}}=
\frac{(n+1)(q-1)}{q^{1+n-\dim V}}\sum\limits_{k=0}^n\frac{ |{\cal
A}_k(V^\perp)|}{k+1}\left(1-\frac{(-1)^{k+1}}{(q-1)^{k+1}}\right).$$
\end{corollary}

A special case of Corollary \ref{corLeont0}  for $q=2$ is proved in
\cite{Leont}.
\begin{equation}\label{eqLeont}
\sum\limits_{k=0}^n\frac{|{\cal A}_k({V})|}{{n\choose
k}}=\frac{n+1}{2^{n-\dim V}}\sum\limits_{k=0\mod 2}\frac{|{\cal
A}_k(V^\perp)|}{k+1}.
\end{equation}

\section{A uniform weight spectrum sequences}

\begin{proposition}\label{unispec2}
Let $V_n$ and $U_n$ be sequences of subspaces of $Q_2^{m_n}$ and
$U_n\subset V_n$. If  $U_n$ has a uniform weight spectrum then $V_n$
has a uniform weight spectrum.
\end{proposition}
Proof. Consider a bipartite graph $G$ with parts $D_1={\cal
A}_i(V_n\oplus w)$ and $D_2=({\cal A}_i(V_n\oplus w)\oplus
U_n)\setminus D_1$ for some weight $i$ and a vector $w$. Without
lost of generality, let $w=\overline{0}$. Vertices $v_1\in D_1$ and
$v_2\in D_2$ are adjacent if and only if $v_2=v_1\oplus u$ where
$u\in U_n$. The degree of  $v\in D_1$ in $G$ is not less than
$|U_n|(1-\alpha(U_n))$. Indeed if $v\oplus u\in D_1$ then ${\rm
wt}(v\oplus u)=i$ and, consequently, $u\in{\cal A}_i(U_n\oplus v)$.
By the definition of a uniform weight spectrum we obtain that
$|{\cal A}_i(U_n\oplus v)|\leq \alpha(U_n)|U_n|$. In the same way we
can prove that the degree of $v\in D_2$ in $G$ is not greater than
$\alpha(U_n)|U_n|$. By double  counting edges we obtain that
$\alpha(U_n)|D_2|\geq |D_1||U_n|(1-\alpha(U_n))$. Then
$$\frac{|{\cal A}_i(V_n)|}{|V_n|}\leq \frac{|D_1|}{|D_2|}\leq
\frac{\alpha(U_n)}{1-\alpha(U_n)}.$$ By the hypothesis of the
proposition $U_n$ has a uniform weight spectrum. Then $V_n$ has a
uniform weight spectrum too. $\square$

It is clear that a sequence of spaces $Q^{m_n}_2$ with growing
dimension $m_n$ has a uniform weight spectrum. Consequently, by
Proposition \ref{unispec2} we obtain the following

\begin{corollary}\label{unispec3}
If  $\mathcal{A}_1(V_n)\rightarrow \infty$  then  a sequence $V_n$
has a uniform weight spectrum.
\end{corollary}

Below we will show that a similar proposition for vectors with
weight $2$ is false.

Now we consider $t$-weight linear and affine codes.
 It is well known the following
\begin{theorem}[Delsarte]\label{thDelsarte}
If $f:Q^n_q\rightarrow \mathbb{R}$ is a function such that
$\widehat{f}(0)\neq 0$ and ${\rm supp}(\widehat{f})\subseteq {\cal
A}_{i_1}\cup\dots\cup{\cal A}_{i_k}$, then $|{\rm supp}({f})|\geq
q^n/b(n,q,k)$ where $b(n,q,k)=|{\cal A}_{0}\cup\dots\cup{\cal
A}_{k}|$.
\end{theorem}

  We obtain the following

\begin{corollary}\label{corDelsarte}
If $C\subset Q^n_q$ is a linear code and $C\subseteq {\cal
A}_{0}\cup{\cal A}_{i_1}\cup\dots\cup{\cal A}_{i_k}$ then $|C|\leq
b(n,q,k)$.
\end{corollary}
Proof. Suppose that $f$ is the indicator function of dual code
$C^\perp$. It is easy to see that ${\rm supp}(\widehat{f})=C$.  The
required statement follows from the equation $|C||C^\perp|=q^n$.
$\square$

Introduce the notation $L(n,q,i_1,\dots,i_k)=\min |{\rm supp}(f)|$,
where ${\rm supp}(\widehat{f})\subseteq {\cal
A}_{i_1}\cup\dots\cup{\cal A}_{i_k}$.

The values of $L(n,q,i,i+1,\dots,j)$ for $q\geq 4$ and $q=3$,
$i+j\leq n$ are calculated in \cite{VV}.  The values of $L(n,2,k)$
were known early from \cite{Krotov}, $L(n,2,k)=
2^{(n+|\theta_k|)/2}$ where $\theta_k=n-2k$.

\begin{corollary}\label{corAf1}
If $C\subset Q^n_q$ is an affine code and $C\subseteq {\cal
A}_{i_1}\cup\dots\cup{\cal A}_{i_k}$ then $|C|\leq
q^n/L(n,q,i_1,\dots,i_k)$.
\end{corollary}
Proof. By the definition of the Fourier transform for any affine
code $C$ the equations $|C||{\rm supp}(\widehat{{\bf1}_C})|=q^n$ and
${\rm supp}\widehat{(\widehat{{\bf1}_C})}=C$ are true. $\square$

\begin{corollary}\label{KrEigen}
If $C\subset Q^n_2$ is an affine code and $C\subseteq {\cal A}_{k}$
then $|C|\leq 2^{(n-|\theta_k|)/2}$.
\end{corollary}

Examples of affine codes that achieve the bound from Corollary
\ref{KrEigen} are $$M_{n,i}=\{(x,x\oplus \bar 1,\bar 0)  : x\in
Q_2^{(n-i)/2}\}$$ where $i=n-2k$.

Let $n$ be even. Consider the linear span of the affine space
$M_{n,0}$. It is equal to $C_n=M_{n,0}\cup (v+M_{n,0})$ where $v\in
M_{n,0}$. It is easy to see that $\dim C_n=1+\frac{n}{2}$ and
$\mathcal{A}_{n/2}(C_n)=2^{n/2}$. So, the sequence $C_n$ does not
get a uniform weight spectrum. Moreover, $C_n$ contains a subspace
$\{(x,x) :x\in Q^{n/2}_2\}$. Consequently, $\mathcal{A}_{2}(C_n)=n/2
\rightarrow\infty$.

\section{Eigenfunctions of the Fourier transform}

The codes $M_{n,0}$ are  extremal in  another sense. Consider
eigenfunctions of the Fourier transform. It is well known that
eigenvalues of the Fourier transform on $Q^n_2$ are equal to $\pm
1$. Known examples of eigenfunctions of the Fourier transform are
functions of type $(-1)^b$ where $b$ is a self-dual bent function.
Functions of type $(-1)^f$ have maximum supports. Let us find
eigenfunctions with a minimum support. A proof of the following
equation can be found in \cite{Tao}.
 \begin{equation}\label{ebent1}
 |{\rm
supp}(f)|\cdot|{\rm supp}(\widehat{f})|\geq 2^n.
\end{equation}

Therefore we obtain that  $|{\rm supp}(f)|\geq 2^{n/2}$ if
$\widehat{f}=\pm f$.

 For even $n$
consider the function $g:Q_2^{n}\rightarrow \{0,\pm1\}$ defined as
\[
g(y)= \left\{
\begin{array}{rr}
(-1)^{{\rm wt}(x)}, & \mbox{ if}\  y=(x,x\oplus \bar 1);\\
0, & \mbox{ otherwise.}\
\end{array}
\right.
\]

It is not difficult to calculate that $g=\widehat{g}$, ${\rm
supp}(g)\subset {\cal A}_{n/2}$ and $|{\rm supp}(g)|=2^{n/2}$.

\section{Covering arrays}

A covering array $C$ with strength $t$ is a subset of $Q^n_q$ having
the following property. For
 any $v\in Q^n_q$ and for each set of $t$ coordinate positions
  there exists  $u\in C$ such that $v$
and $u$ can differ from each other only in these $t$ positions.
I.e., the $t$-dimensional face of the hypercube defined by these
positions intersects $C$. The problem of the construction of
covering arrays with a minimum cardinality is very popular (see
\cite{LKLK}). We consider a little different problem. It is
necessary  to find a subset (partial covering array) $S\subset
Q^n_2$ having cardinality $2^{n-t}$ and intersecting with maximum
 number of $t$-dimensional faces. In \cite{Pot} it is discussed an application of partial
covering arrays for videocoding and cryptography. By linear
algebraic arguments we can verify the following

\begin{proposition}\label{pvs111} Let $C\subset Q^n_2$ be an affine $(n-k)$-dimensional space
  and  $\Gamma$ be a $k$-dimensional face.
Then $|C\cap (x+\Gamma)|=0\ {\rm or}\ 2^s$ for each $x\in Q_2^n$,
where $s$ does not depend on $x$.
\end{proposition}

\begin{corollary}\label{corAf5}
The number of $\frac{n}{2}$-dimensional faces $\Gamma\subset Q^n_2$
such that $\Gamma\cap M_{n,0}\neq \varnothing$ is greater than
$2^n$.
\end{corollary}
Proof. There is a one-to-one correspondence between elements of
$M_{n,0}$ and $\frac{n}{2}$-dimensional faces $\Gamma$ such that
$\bar 0\in \Gamma$ and $\Gamma\cap M_{n,0}\neq \varnothing$.
Consequently, we have $2^{n/2}$ such faces. By Proposition
\ref{pvs111} for any such face $\Gamma_0$  we obtain  $2^{n/2}$
different faces $x+\Gamma_0$ such that $(x+\Gamma_0)\cap M_{n,0}\neq
\varnothing$. $\square$

We can suggest that $M_{n,0}$ intersects the maximum number of
$\frac{n}{2}$-dimensional faces among all sets with cardinality
$2^{n/2}$.


\begin{thebibliography}{9}




\bibitem {MS}
F.J. MacWilliams and N.J.A. Sloane, \emph{The Theory of
Error-Correcting Codes}. Elsevier/North-Holland, Amsterdam, 1977.



\bibitem {Pless}
V. Pless,  ``Power moment identities on weight distributions in
error correcting codes'', \emph{ Information and Control}, vol.6,
(1963), pp. 147--152.




\bibitem {Leont}
V.K. Leont'ev, ``On spectra of linear codes'', \emph{ Probl.
Peredachi Inf.}, vol.53(4), (2017), pp.  343--348.


\bibitem {Sid}
V.M. Sidelnikov, \emph{Teoriya kodirovaniya}. Fizmatlit, Moscow,
2008 (in Russian).

\bibitem {Krotov}
D.S. Krotov, ``Traids in the combinatorial configurations'', XII
International Seminar "Discrete Mathematics and its Applications"
(Moscow, 20-25 June 2016), pp. 84--96 (in Russian).

\bibitem {VV}
A. Valyuzhenich and K. Vorob'ev,  ``Minimum supports of functions on
the Hamming graphs with spectral constraints'', \emph{  Discrete
Math.}, vol.342(5), 2019, pp. 1351--1360.


\bibitem{Tao}
T. Tao,  ``An uncertainty principle for cyclic groups of prime
order'', \emph{Math. Res. Lett.}, vol.12(1), 2005, pp. 121--127.


\bibitem{Delsarte}
P. Delsarte, ``Four fundamental parameters of a code and their
combinatorial significance'', \emph{ Information and Control},
vol.23, 1973, pp. 407--438.

\bibitem{Linial}
N. Linial and J. Mosheiff,  ``On the weight distribution of random
binary linear codes'', \emph{  Random Structures Algorithms},
vol.56(1), 2020, pp. 5--36.

\bibitem {LKLK}
 J. Lawrence,  R.N. Kacker, Y. Lei, D.R. Kuhn,  M. Forbes,  ``A survey of
binary covering arrays'', \emph{  Electronic J. of Combinatorics},
vol.18(1), 2011.

\bibitem{Pot}
V.N. Potapov,  ``Partial covering arrays for data hiding and
quantization'', \emph{ Siberian Electronic Mathematical Reports},
vol.15, 2018, pp. 561--569.




\end{thebibliography}
\end{document}